\documentstyle[prl,aps,epsf]{revtex}
\draft
\begin{document}
\twocolumn[\hsize\textwidth\columnwidth\hsize\csname @twocolumnfalse\endcsname

\title{Moving Wigner Glasses and Smectics: Dynamics of Disordered Wigner Crystals}
\author{C. Reichhardt$^{(1)}$, C.J. Olson$^{(1)}$, and Franco Nori$^{(2)}$}
\address{{\rm (1)} 
Department of Physics, University of California, 
Davis, California 95616}
\address{{\rm (2)} Department of Physics, 
University of Michigan, Ann Arbor, MI 48109-1120}  

\date{\today}
\maketitle
\begin{abstract}
We examine the dynamics of driven classical Wigner solids 
interacting with quenched disorder from charged impurities.
For strong disorder, the initial motion is plastic---in
the form of crossing winding channels.
For increasing drive, the disordered Wigner glass can 
reorder to a moving Wigner smectic--with the electrons 
moving in non-crossing 1D channels. These different 
dynamic phases can be related to the conduction noise
and $I(V)$ curves.
For strong disorder, we show criticality in the 
voltage onset just above depinning. 
We also obtain the dynamic phase diagram for driven Wigner 
solids and prove that there is a finite threshold for 
transverse sliding, recently found experimentally.   
\end{abstract}

\vspace*{-0.4in}
\vskip2pc]
\narrowtext

Ordered arrays of charged particles 
have been studied in the context of
colloidal suspensions, ion rings,
atomic-ion Wigner crystals, quantum computers, 
astrophysics, biophysics, plasmas,
electrons deposited on liquid Helium surfaces, 
semiconductor heterostructures, Wigner droplets, 
and arrays of metallic islands interconnected 
by tunnel junctions \cite{overview}.
A revival of general interest in charged 
arrays has been fueled by the observation, 
in 2D heterostructures, of  
nonlinear I-V curves 
exhibiting a threshold as a function of an 
externally applied electric field 
\cite{Goldman,Pinned},
indicating the presence of a Wigner solid (WS) \cite{WC} 
that has been pinned by the disorder in the sample. 
The depinning threshold 
can
vary up to two orders of magnitude
in different samples \cite{Pinned}. 
Indeed, experiments \cite{overview} 
based on transport 
and photoluminescence 
provide indirect evidence (without observing 
the location of the charges) for the existence 
of the WS, and demonstrate 
the very important role {\it disorder\/} plays 
in the dynamics of the WS \cite{disorderedWC}. 
It is the purpose of this work to study how disorder 
affects the transport properties of the driven WS.

The onset of broadband conduction noise has been interpreted
as a signature of the sliding of a defected WS
 \cite{Goldman}.
If the electrons were to retain their order and slide collectively,
narrow band noise resembling that seen in 
sliding charge-density wave systems 
should be observed instead.
Simulations by Cha and Fertig on classical Wigner 
crystals interacting with charged defects 
indicate that a disorder-induced transition, 
from a clean to a defected WS, can occur as 
a function of increasing pinning strength \cite{Fertig}. 
For strong pinning, the initial depinning is plastic and 
involves tearing of the electron crystal \cite{Fertig}(b).
However, many aspects of this transport have not been 
systematically characterized, including the current-voltage 
$I(V)$ curves, conduction noise, transverse meandering, 
transverse threshold for sliding, and the electron lattice 
structure and motion for varying applied drives. 
 
Similar plastic depinning transitions have been observed in the
related system of driven vortex lattices in disordered superconductors.
The plastic depinning of the vortices is associated with
flux motion through intricate river-like channels \cite{Jensen,Olson}.
Defects in the vortex lattice strongly affect 
the depinning thresholds 
and the voltage noise 
signatures produced by the system. 
For increasing drive the
initially defected vortex lattice can reorder to 
a moving lattice or a moving smectic phase 
\cite{Koshelev,Giamarchi,Balents}.
In the fast moving lattice phase 
the vortex lattice regains order in both transverse
and longitudinal directions with respect to the driving force, 
while in the smectic state 
only order transverse to the driving direction appears.       

It is unclear a priori whether the same type of 
reordering transitions can occur in a system 
containing {\it long-range\/} pinning, such as 
in the case of the WS interacting with charged 
impurities.  This is in contrast with the vortex 
system, where the pinning interacts with the 
flux lines only on a very short length scale.  
Additionally, in the strong pinning limit, critical 
behavior may occur near the depinning threshold, 
leading to velocity-force relations \cite{Fisher} 
of the form $v \sim (F - F_{T})^{\xi}$. 
For instance, for transport in metallic dots,
$\xi = 5/3$ theoretically \cite{middleton}, and 
$ \xi = 2.0$ and $1.58$ experimentally \cite{Kurdak}.   

In this work, we use numerical simulations to study dynamical 
transitions in a 2D electron system forming a classical WS in 
the presence of charged impurities. 
For strong pinning, the electron crystal is highly defected and 
depins plastically, with certain electrons flowing in well-defined
channels or interconnecting rivers, while others remain immobile. 
For increasing applied driving force, the initially disordered  
electrons can partially {\it reorder} to a moving Wigner smectic 
state where the electrons flow in 1D non-crossing 
winding channels. 
The reordering transition is accompanied by a
saturation in the $dI/dV$ curves 
and by a change from a broad-band to
a narrow-band voltage noise signature. 
For weak disorder, the depinning is elastic
and a narrow-band noise signal can be observed
at all drives above depinning.  
For strong disorder, where the depinning 
is plastic, we find criticality in the
velocity force curves in agreement with 
transport in metallic dots \cite{middleton,Kurdak}. 
We map out the dynamic phase diagram as a function 
of disorder strength and applied driving force.
Also, we find unequivocal evidence for a transverse 
finite threshold for sliding conduction, in agreement 
with recent experiments \cite{transverse}.
Our results can also be tested in other systems listed 
in the first sentence of this paper (e.g., with charged 
colloidal systems suspended in water in the presence 
of a disordered substrate).

We conduct overdamped molecular dynamics (MD) simulations 
using a model similar to that studied by Cha and Fertig 
\cite{Fertig}.  The energy from interactions is 
\vspace*{-0.1in}
\begin{equation}
U = \sum_{i \neq j}\frac{e^{2}}{|{\bf r}_{i} - {\bf r}_{j}|} 
- \sum_{ij}\frac{e^{2}}{\sqrt{(|{\bf r}_{i} - {\bf r}^{(p)}_{j}|^{2} + d^{2})}} 
\vspace*{-0.1in}
\end{equation}
The first term is the electron-electron (Coulomb) repulsion and
the second term is the electron-impurity interaction, 
where the impurities are positively charged defects out of plane. 
Here,
${\bf r}_{i}$ is the location of electron $i$, 
and ${\bf r}_{j}^{(p)}$ is the in-plane location 
of a positive impurity which is located at
an out-of-plane distance $d$ (measured here in 
units of $a_0$, the average lattice constant of 
the WS).  The number of electrons $N_i$ 
equals the number of impurities $N_p$, and the 
{\it disorder strength\/} is varied by changing $d$.
We have also considered many other cases (e.g., 
$\pm$ charged impurities, $N_i \neq N_p$, etc.) 
with consistent results.  
The long-range Coulomb interactions are evaluated with 
a fast converging sum technique by Jensen \cite{niels} 
which is computationally more efficient than the
Ewald-sum technique used in previous studies \cite{Fertig}. 

To obtain the initial electron positions, we perform simulated
annealing in which we start from a high temperature
above the electron lattice melting transition and slowly
cool to a low temperature. Temperature is implemented via 
Langevin random kicks.  Once the electron configuration 
has been initialized, the critical depinning force is determined 
by applying a very slowly increasing uniform driving force
$f_{d}$ which would correspond to an applied electric field.
After each drive increment, we wait $10^{4}$ MD time steps 
before taking data, which we average over the next $10^{4}$ 
time steps.   For each drive increment, we measure the 
average electron velocity ($\propto$ current) 
in the direction of drive, 
$V_{x} = (1/N_i) \sum_{i}^{N_{i}}{\hat {\bf x}}\cdot {\bf v}_{i}$.  
The  $V_{x}$ versus $f_{d}$ curve corresponds to an $I(V)$ 
experimental curve and we will thus use the notation $V_x=I$ and $f_d=V$.
The depinning force $f_{d}^{c}$ is defined 
as the drive $f_{d}$ at which $V_{x}$ reaches a value 
of $0.01$ that of the ohmic response. 
We have studied system sizes $N_{i}$ 
from $64$ to $800$ and find similar behavior at all sizes.  
Most of the results presented here are for systems with 
$N_{i}=256$.  

\begin{figure}
\center{
\epsfxsize=3.5in
\epsfbox{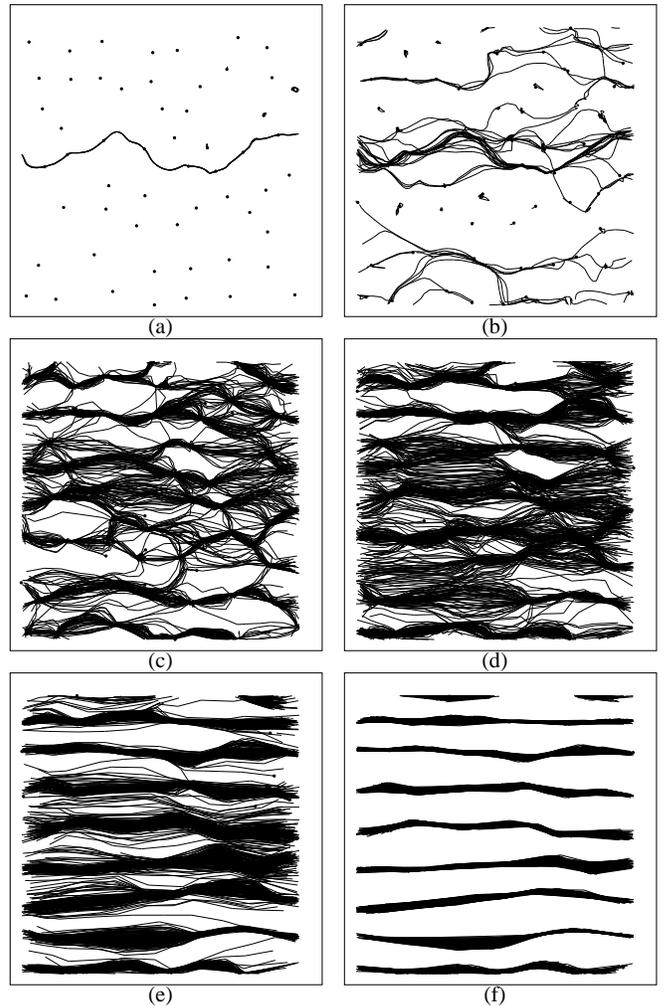}}
\caption{Electron positions (dots) and eastbound trajectories (lines), 
for a sample with $N_i/N_p=1.0$ and $d=0.65$.
$f_d/f_{d}^{c} =$ (a) 1.1, (b) 1.5, (c) 2.25, (d) 3.0, (e) 4.0, (f) 5.0.}
\label{fig:image}
\end{figure}

When driven through a sample containing strong pinning, 
the electron lattice undergoes a 
gradual 
reordering transition as the driving force is increased.
We illustrate such a reordering transition in Fig.~\ref{fig:image}
by plotting the eastbound 
electron trajectories at different driving forces for 
a sample containing strong disorder of $d=0.65$.
In Fig.~\ref{fig:image}(a) the onset of motion occurs through 
the opening of {\it a single\/} winding channel.  Electrons outside 
of the channel remain {\it pinned}, and the overall 
electron lattice is disordered.  At $f_{d}/f_{d}^{c} = 1.5$, 
shown in Fig.~\ref{fig:image}(b), 
several channels 
have opened, some of which are interconnecting.

\hspace{-15pt}
The original channel in Fig.~\ref{fig:image}(a) has grown in 
width, but regions of pinned electrons are still present.  
Electrons moving past a pinned electron perturb it, causing 
it to move (like a revolving door) in a circular orbit around 
the center of the potential minima in which it is trapped. 
Several of these {\it electron turnstiles\/} can be observed 
in Fig.~\ref{fig:image}(b).
In Fig.~\ref{fig:image}(c) for $f_{d}/f_{d}^{c} = 2.25$ there
appear to be regions where electrons do not flow, but none of
the electrons are permanently pinned.  Some electrons 
become pinned for a period of time before moving again.
If the trajectories are drawn for a sufficiently long time, 
the electron flow appears everywhere in the sample, although 
there are still preferred paths in which more electrons flow. 
In Fig.~\ref{fig:image}(d) for $f_{d}/f_{d}^{c} = 3.0$ the 
electron flow occurs more uniformly across the
sample. In Fig.~\ref{fig:image}(e) at $f_{d}/f_{d}^{c} = 4.0$ 
the electrons begin to flow predominantly in certain 
non-crossing channels, although some electrons jump
from channel to channel. In Fig.~\ref{fig:image}(f) 
for $f_{d}/f_{d}^{c} = 5.0$ the electron 
flow occurs in well defined non-

\begin{figure}
\center{
\epsfxsize=3.5in
\epsfbox{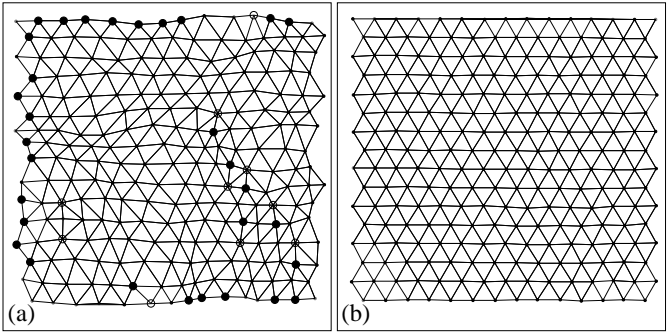}}
\caption{Delaunay triangulation for electrons in a sample 
with (a) $d=0.65$ (strong pinning) at $f_d/f_{d}^{c}=5.0$; and 
(b) $d=1.76$ (weak pinning) and $f_d/f_{d}^{c}=2.0$.  Large 
circles indicate (5- or 7-fold) defects in the electron lattice.}
\label{fig:delaunay}
\end{figure}

\hspace{-15pt}
crossing channels, 
which can contain different numbers of electrons.
A similar channel motion exists for driven vortices 
in disordered superconductors \cite{Giamarchi,Balents,Moon,Olson}.

For samples containing very weak disorder, the pinned WS has 
six-fold ordering and depins elastically, without generating 
defects in the lattice.  In this regime, the electron crystal 
flows in 1D channels with each channel containing the {\it same} 
number of electrons. Here, the transverse wanderings of the 
electrons are 
considerably reduced compared to the case of strong pinning. 

To better illustrate the change in the amount of disorder in the
electron lattice,
in Fig.~\ref{fig:delaunay}(a) 
we show the Delaunay triangulation for the electrons 
for the same drive as in Fig.~\ref{fig:image}(f). 
Defects, in the form of 5-7 disclination pairs, appear with their
Burgers vectors oriented perpendicular to the direction of the drive.
In the structure factor there are only two prominent peaks
for order in the direction transverse to the drive, consistent 
with a moving Wigner smectic state. 
In Fig.~\ref{fig:delaunay}(b) we show the Delaunay triangulation
for the moving state 
in a weakly pinned sample 
with $d = 1.76$ where the initial depinning is elastic. 
Here the moving lattice is defect free.
The structure factor in this case shows four
longitudinal peaks in addition to the two transverse peaks,
although the transverse peaks are more prominent. 
Much larger systems would be necessary to determine whether the 
system is in a smectic state or in a moving Bragg glass
in this case of weak disorder.
 
\begin{figure}
\center{
\epsfxsize=3.5in
\epsfbox{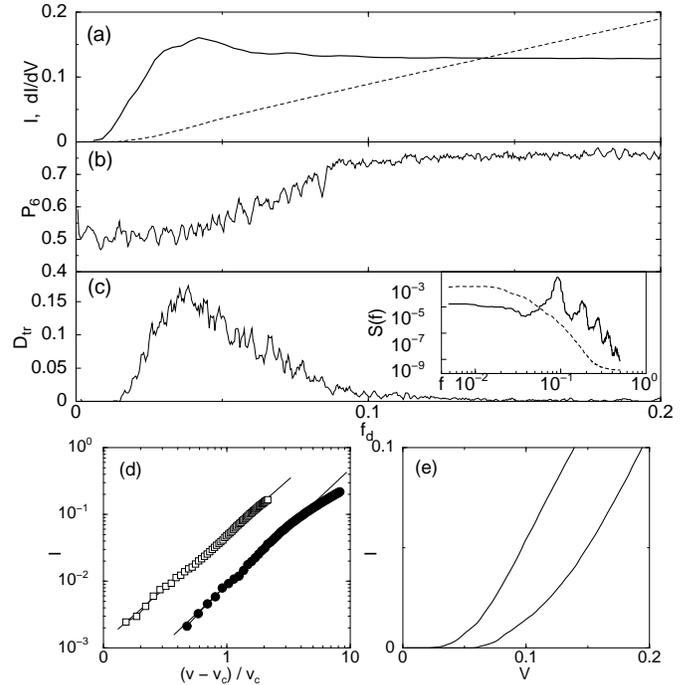}}
\caption{
(a) $I(V)$ (dashed line) and $dI/dV$ curves (solid line)
for a sample with $d=0.65$.
(b) Fraction $P_6$ of sixfold coordinated electrons, as a function
of driving force $f_d$.  (c) Fraction $D_{tr}$ of 
transversely wandering electrons.
Inset to (c): Noise spectra for $f_{d}/f_{d}^{c}=1.5$ in the
plastic flow regime showing broad-band noise (dashed line) and
for $f_{d}/f_{d}^{c}=4.0$ in the smectic regime 
showing a narrow-band signal.
(d) Log-log plot of
$V_x = (f_{d}-f_{d}^{c})^{\xi}$, in which $\xi= 1.61 \pm 0.10$ and
$1.71$.  (e) $V_x$ versus $f_d$ 
for disorder strengths of $d=0.5$ and $0.65$.
}
\label{fig:iv}
\end{figure}

In order to connect the reordering sequence with a measure 
that is readily accessible experimentally, we plot in 
Fig.~\ref{fig:iv}(a) the $I(V)$ and $dI/dV$ curves. 
A peak occurs in  $dI/dV$ at $f_{d}/f_{d}^c \approx 3$, when
the electrons are undergoing very disordered plastic flow.
In Fig.~\ref{fig:iv}(b) we plot the fraction $P_{6}$ of
six-fold coordinated electrons as a function of drive.
A perfect triangular lattice would have $P_{6} = 1$. 
For drives 
$f_{d}/f_{d}^{c} < 2$ the lattice is highly defected.
For $f_{d}/f_{d}^{c} > 3$ (i.e., past the peak in $dI/dV$), 
the order in the lattice begins to increase.
The value of $P_6$ saturates near $f_{d}/f_{d}^{c} \approx 6$  
which also coincides with the saturation 
of the $dI/dV$ curve.  Thus the 
experimentally observable $I(V)$ 
characteristic can be considered a good measure of 

\hspace{-15pt}
the {\it degree\/} of order and the {\it nature\/} 
of the flow in the system.  
To quantify the {\it degree of plasticity\/} of the 
electron flow, we plot in Fig.~\ref{fig:iv}(c) the 
fraction of electrons $D_{tr}$ that wandered a 
distance of more than $a_{0}/2$ 
in the direction {\it transverse\/} to the drive
during an interval of 8000 MD steps.
The peak in $D_{tr}$ coincides with the peak in $dI/dV$. 
$D_{tr}$ then slowly declines until it saturates 
at $f_{d}/f_{d}^{c} \approx 6$, indicating the {\it gradual\/}
formation of the {\it non-crossing channels\/} as seen in 
Fig.~\ref{fig:image}(f). The saturation in $D_{tr}$ 
coincides with the saturations in both $P_{6}$ and $dI/dV$.    

The $I(V)$ curve in Fig.~\ref{fig:iv}(a) corresponds 
to a highly irregular voltage signal as a function 
of time when the electrons are in the plastic 
flow regime ($f_{d}/f_{c} = 1.5$).
The corresponding voltage noise spectrum 
in the inset of Fig.~\ref{fig:iv}(c) 
shows that only
broad-band noise is present.  
In contrast, at $f_{d}/f_{d}^{c} = 4.0$, 
in the moving smectic regime,
a roughly regular signal is obtained, and 
narrow-band noise 
appears, as shown in the inset
of Fig.~\ref{fig:iv}(c).
In a system with $d = 1.57$ when the depinning is elastic only,
an even more pronounced narrow band noise signal is observable. 
In experiments, broad band noise has been observed above depinning 
\cite{Goldman}, but narrow band noise has not been seen.
Our results suggest that
{\it plasticity} may be playing an
important role in most experiments.  

\begin{figure}
\center{
\epsfxsize=3.5in
\epsfbox{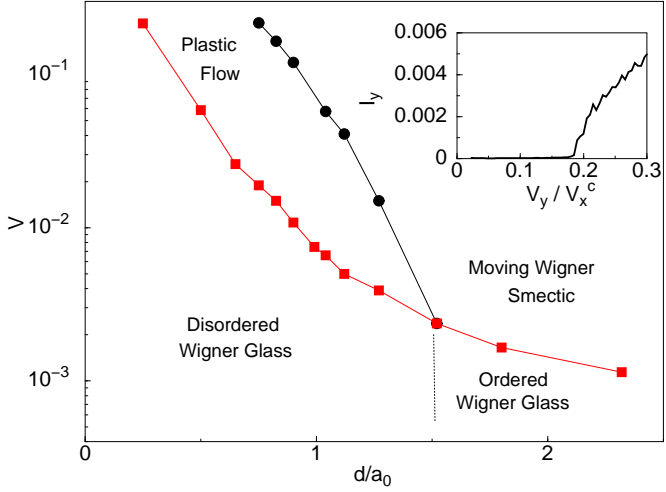}}
\caption{Dynamic phase diagram for the driven disordered Wigner solid.  
Inset: Clear evidence for a finite transverse depinning threshold
for a system with $d=0.9$ at $f_d=0.16$ in the reordered phase.  
}
\label{fig:phase}
\end{figure}

\hspace{-18pt}

In Fig.~\ref{fig:iv}(d,e) we examine the critical behavior 
above the depinning threshold of $V_x$ versus 
$f_d$ for disorder strengths 
$d= 0.5$ and $0.65$. In Fig.~\ref{fig:iv}(d), 
a log-log 
plot of $V_x \sim [(f_{d} -f_{d}^{c})/f_d^{c}]^{\xi}$, 
indicates that the curves are fit well by a power law 
over one and half decades with 
$\xi = 1.61 \pm 0.10$ and $1.71 \pm 0.10$,
respectively. These values 
agree well with the predicted value of $\xi = 5/3$ \cite{middleton} 
for electron flow through disordered arrays, and the
experimentally observed values of  $\xi = 1.58$ and $2.0$ 
\cite{Kurdak}. 

The theory of the moving ordered phase \cite{Giamarchi}
predicts 
a barrier to transverse motion
once channels similar to those in Fig. 1(e) form.
As shown in the inset of Fig. 4, for a system in the
reordered phase we observe a
transverse depinning threshold that is about 1/6 
the size of the longitudinal depinning threshold
$f_d^{c}$.  
In recent experiments, Perruchot {\it et al.} \cite{transverse} 
also find evidence for a transverse barrier that is
about 1/10 the size of the longitudinal threshold.  
These thresholds are much larger than those observed in vortex matter
interacting with short-ranged pinning, where ratios of 1/100 
are seen \cite{Moon}.

In Fig.~\ref{fig:phase} we present the WS {\it dynamic phase diagram\/}
as a function of disorder strength and driving force.   
For $d < 1.4$ there are a considerable number of defects 
in the WS, and the initial depinning is plastic.
We label this region the {\it pinned Wigner glass}.
For increasing disorder, the pinned region grows while 
for strong enough drive the electrons reorder to a 
{\it moving Wigner smectic\/} state.  For $d > 1.4$ 
there are few or no defects in the pinned state and the
initial depinning is elastic.  
The presence of a crossover from elastic to plastic depinning
with decreasing disorder strength is in agreement  
with 
\cite{Fertig}.  

In summary, we have investigated the pinning and dynamics of an
electron solid interacting with charged disorder. We find that 
for strong disorder the depinning transition is plastic with 
electrons flowing in a network of winding channels. For 
increasing drives, the electrons partially reorder and 
flow in non-crossing channels forming a moving Wigner smectic. 
We show that the onset of these different phases can be 
inferred from the transport characteristics.  
In the plastic flow regime, the noise has broad-band 
characteristics, while in the moving smectic or elastic 
flow phase, a narrow band noise signal is observable.  
We also show that the onset of the plastic flow phase 
shows critical behavior with critical exponents in 
agreement with predictions for transport in 
arrays of metallic dots \cite{middleton,Kurdak}. 
We map out the 
dynamic phase diagram as a function 
of disorder and applied driving force.  
We 
obtain a finite threshold for transverse sliding, 
in agreement with recent experiments \cite{transverse}.

We gratefully acknowledge J.~Groth, Niels Gr{\o}nbech-Jensen, 
and Alan H. MacDonald for useful discussions, and B. Janko
for his kind hospitality. 
This work was partially supported by DOE Office of Science No.
W-31-109-ENG-38, CLC and CULAR (LANL/UC), and NSF DMR-9985978.

\end{document}